\RequirePackage{fix-cm}
\documentclass[twocolumn]{svjour3}          
\smartqed  
\usepackage[utf8]{inputenc}
\usepackage{graphicx}
\usepackage{mathptmx}      
\usepackage{amsmath}
\usepackage{natbib}
\usepackage{soul}
\usepackage{color}
\journalname{}
\begin{document}

\title{The differences of damage initiation and accumulation of DP steels: a numerical and experimental analysis
}


\author{Felix Pütz *         \and
        Fuhui Shen \and
        Markus Könemann \and
        Sebastian Münstermann 
}


\institute{* Felix Pütz \at
              Integrity of Materials and Structures, Intzestraße 1 52072 Aachen\\
              Tel.: +49-241-8028415\\
              Fax: +49-241-8092253\\
              \email{felix.puetz@iehk.rwth-aachen.de}             \\
}

\date{Received: date / Accepted: date}

\maketitle

\begin{abstract}
Many studies have examined the damage behaviour of dual-phase steels already. It is a topic of high interest, since understanding the mechanisms of damage during forming processes enables the production of steels with improved properties and damage tolerance. However, the focus was rarely on the comparison between representatives of this steel class, and the numerical simulation for the quantification of damage states was not thoroughly used. Therefore, this study compares the damage initiation and accumulation of two dual-phase steels (DP800 and DP1000), which are used in the automotive industry. Additionally, parameter sets of a phenomenological damage mechanics model with coupled damage evolution are calibrated for each material. The combined analysis reveals an earlier initiation of damage for the DP800, where the damage accumulation phase is prolonged. For DP1000 the damage nucleates only shortly before material failure. The material model is able to correctly predict the behaviour, while experimental analysis confirms the prediction via light optical and SEM metallography.
\end{abstract}

\section{Introduction}
\label{intro}
The usage of dual-phase (DP) steels has been on the rise in recent years. Especially the automotive industry shows high interest in the development of these advanced high strength steels (AHSS), since DP steels show high strength values while still maintaining good formability. Thus, a lightweight component design can be achieved by reducing component thickness while still keeping the identical safety conditions \citep{Davies.1979}. These specific properties result from a distinct microstructure, that is composed of a soft ferritic phase with hard martensite islands on the grain boundaries and triple points of ferrite grains. Due to the difference in mechanical properties of the two phases, plastic behaviour of DP steels shows a relatively low yield to tensile ratio, pronounced strain hardening and excellent global formability. The reason for this extraordinary property profile lies in the partitioning of stress and strain between the involved phases, allowing for multiple degrees of freedom for microstructural design \citep{Bieler.2009} . \newline

The strain partitioning between ferrite and martensite depends vastly on the specific microstructure. Marteau et al. reported that the local microstructural neighbourhood is the critical factor for strain heterogeneity \citep{Marteau.2013}. Strain accumulates mostly in the ferrite forming localized bands with an angle of 45-50$^\circ$ with respect to the loading direction \citep{Ghadbeigi.2010, Tasan.2014}, whereas the martensite carries the majority of the applied stress \citep{Tasan.2014b}. Therefore, martensite is elastically deformed for materials with low martensite content, while its deformation behaviour is plastic for high contents \citep{Shen.1986}. The local microstructure especially is determining the strain distribution, e.g. average size of martensite islands and global distribution \citep{Park.2014, Saai.2014}. \newline

Due to this inhomogeneity in the material constituents’ behaviour, the microscopic damage modes of dual-phase steels differ quite significantly to those of common structural steels. Where for structural steels the inclusions play the major role for void initiation, in DP steels damage incidents occur in relation to the two phases, martensite and ferrite \citep{Tasan.2010}. Mechanisms for damage initiation in dual-phase steels are mostly decohesion of the martensite/ferrite interface, cracking of the martensite phase, or a localization of plastic strain in the ferrite phase, which results in debonding of the ferrite grain boundaries \citep{Ahmad.2000}. The mode for the damage initiation depends on the microstructure and the resulting strain heterogeneity \citep{Kadkhodapour.2011}. Therefore, grain size and martensite content do play an important role \citep{Maire.2008, Ramazani.2013, Tasan.2015}. Additionally, martensite morphology influences the early damage nucleation \citep{Ghadbeigi.2013, He.1984}. Besides, observations have shown, that for banded martensite cracking is far more likely than a decohesion of the interface boundary of ferrite and martensite \citep{AvramovicCingara.2009}. \newline

To assess the material's properties and predict the load bearing capabilities of structures, damage mechanics models are widely used for DP steels, e.g. in the automotive industry. In the field of damage mechanics, two different model types exist: Coupled and uncoupled models \citep{Besson.2010}. For the coupled damage mechanics models usually a damage variable is employed to reduce the yield potential according to the softening resulting from ductile damage in the material during deformation. In case of the coupled models, micromechanical models are very popular, for instance the Gurson-Tvergaard-Needleman (GTN) model \citep{Gurson.1977, Tvergaard.1981, Tvergaard.1984}. Micromechanical models are characterized by the depiction of physical phenomena like void nucleation, growth and coalescence through sets of parameters. Therefore, the parameters are interdependent and thus, an extensive iteration process is necessary for the parameter calibration \citep{West.2012}. Alternatively, phenomenological, coupled models are used to describe the damage in materials numerically. In contrast to the micromechanical models, damage evolution is treated in a macroscopic way, where a number of effects are described by a mathematical formulation. A good example for this type of model is the Lemaitre model \citep{Lemaitre.1985, Lemaitre.1992}, which describes damage as an irreversible process.  \newline

Contrary to that, uncoupled models describe the material behaviour including fracture without taking damage into account. Both the Johnson – Cook \citep{Johnson.1985}, as well as the Bai-Wierzbicki (BW) model are good examples for this type of model \citep{Bai.2008}. Further development has been applied by Lian et al., who combined the advantages of uncoupled and coupled models into a hybrid formulation, making it the modified Bai-Wierzbicki model (MBW) \citep{Lian.2012}. The model therefore holds an easy formulation and combines it with the influence of damage onto material behaviour. The model has been developed further since its inception. For the first version a locus for the damage initiation point, which was dependent on both stress triaxiality and Lode angle was utilized. Additionally, a set of critical values for the damage variable was applied, at which material fracture was assumed in the numerical simulation. Wu et al. changed that considerably by implementing a locus for the fracture, as well as considering non proportional loading paths until the inception of ductile damage \citep{Wu.2017}. A further development of the MBW model was made by Shen et al. to characterise the influence of loading orientation, which was used to describe the anisotropic ductile damage and fracture behaviour of pipeline steels \citep{Shen.2020}.  Since the MBW damage mechanics model is easy to use and calibrate, while also depicting the damage behaviour accurately, it is applied here for the characterisation of  damage behaviour in DP steels. \newline

While many studies focused on the damage in dual-phase steels from an experimental standpoint, it is hard to experimentally determine the evolution of damage during the tests. Therefore, this study aims to enhance the experimental investigation by performing finite element (FE) based numerical simulations that are utilized to quantify the damage in the material during forming processes. Thus, in this study, two dual-phase steels, DP800 and DP1000 were compared. Their damage and fracture properties are distinctly different, while the strength is not very far apart. To compare the materials behaviour, a damage mechanics model has been used that can describe both, damage initiation as well as ductile material fracture, while also taking the changes of the stress state during deformation, due to non-proportional loading effects, into account . This allows a comparison of the damage initiation for different stress states between the materials. Additionally, by means of a calibrated fracture locus, the damage accumulation phase can be analysed and compared. Thus, tensile tests  were conducted on flat specimens of different geometries to gather information about materials deformations and damage properties under different stress states. On that basis, the material parameters of the modified Bai Wierzbicki model were calibrated. For the validation of the numerical results regarding the damage initiation and accumulation of the investigated material, interrupted tensile tests were conducted and a metallography analysis was performed by using the light optical microscopy and scanning electron microscopy (SEM).

\section{Materials Characterization}
\label{MatChar}
In the present study, two dual-phase steels were evaluated for comparison purpose. Even though both materials are dual-phase steels, vastly different properties are observable. These varying characteristics are obtained by distinct alloying concepts as well as heat treatment processes.
\begin{figure*} [!h]
  \includegraphics[width=\textwidth]{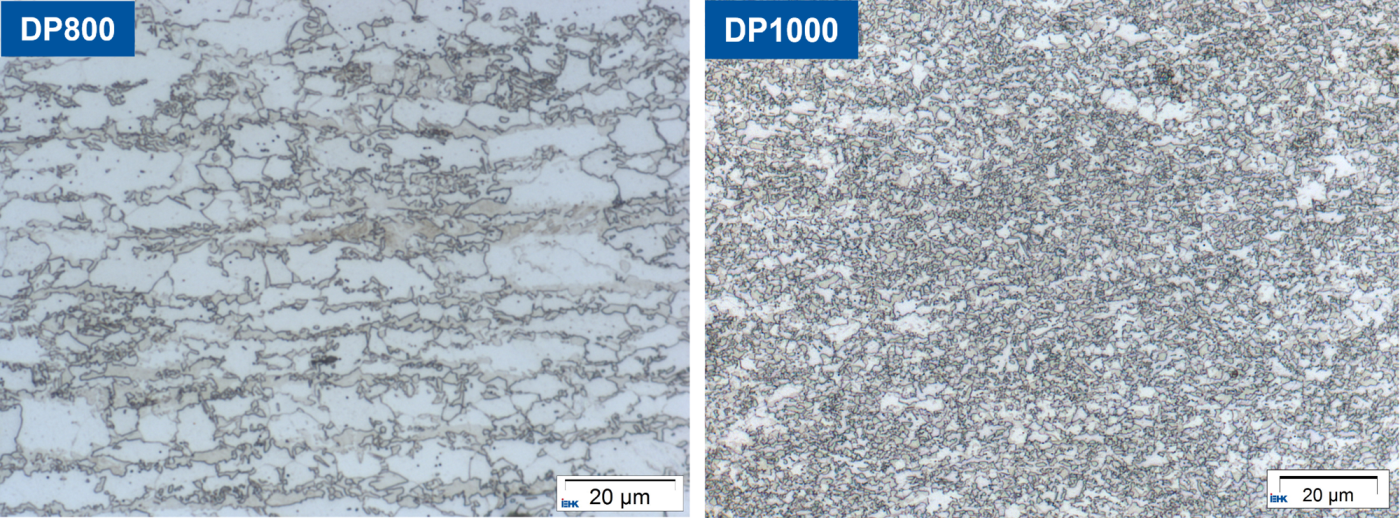}
\caption{Microstructures of steels DP800 and DP100 revealed by HNO3 etching, in light optical metallography}
\label{fig:1}       
\end{figure*}

 \textbf{Fig. \ref{fig:1}} shows a comparison of the respective microstructures at a magnification of 1000. It is very well observable that the average grain size of DP1000 is significantly smaller than that of DP800. Additionally, DP1000 has increased martensite contents of approximately 38\% while DP800 contains about 32\%. For DP800 a pronounced banding of the martensite in the microstructure is noticeable. Since the martensite bands run parallel to the rolling direction, there will be a significant influence on the mechanical properties.  A certain extent of failure anisotropy is expected due to the banded microstructure, however, the anisotropic fracture properties are beyond the scope of this study and all tensile specimens were manufactured perpendicular to the rolling direction of both DP steels.
Both steels were delivered with a thickness of 1.5mm; their respective chemical compositions are given in \textbf{Table \ref{tab:1}}. While the alloying concepts show noticeable similarities, some minor differences are present.

\begin{table}
\caption{Chemical composition of dual-phase steels DP800 and DP1000, in mass-\%}
\label{tab:1}       
\begin{tabular}{lllllllll}
\hline\noalign{\smallskip}
 & C & Si & Mn &  Cr & Mo & Cu  \\
\noalign{\smallskip}\hline\noalign{\smallskip}
\textbf{DP800} & 0.15 & 0.21 & 1.67 &  0.73 & 0.01 & 0.044 \\
\textbf{DP1000} & 0.14 & 0.32 & 1.97 & 0.40 & 0.05 & 0.023 \\
\noalign{\smallskip}\hline
\end{tabular}
\end{table}

The carbon content for DP1000 is decreased compared to DP800, thus leading to higher carbon concentration in the martensite phase for the DP800, since the phase fraction of martensite is higher for DP1000. Thus, it is to be expected that the strength of the martensite is reduced for DP1000 due to the decreased carbon content, therefore leading to a bigger contrast of properties between ferrite and martensite in the DP800.  Furthermore, manganese and chromium contents are different, which leads to slight disparities due to solid solution hardening. Additionally, Si as well as Mn and Cr reduce the critical cooling rate needed for forming martensite, thus influencing the respective time - temperature - transformation graphs. On top of that the solubility of carbon in ferrite is reduced by silicon. Therefore, both materials will have very distinct processing routes tailored to the respective production process. For the characterization of the resulting mechanical properties, isothermal, uniaxial quasistatic tensile tests were carried out on flat specimens without a notch. To ensure a proper depiction of the material's properties, three tests were carried out per DP steel. A video extensometer was used to capture the elongation of the specimen during deformation, where the starting length of 40mm was used. The necking took place inside the area tracked by the extensometer for all 6 tensile tests, ensuring a good comparability.

\begin{figure}
   \includegraphics[width=0.5\textwidth]{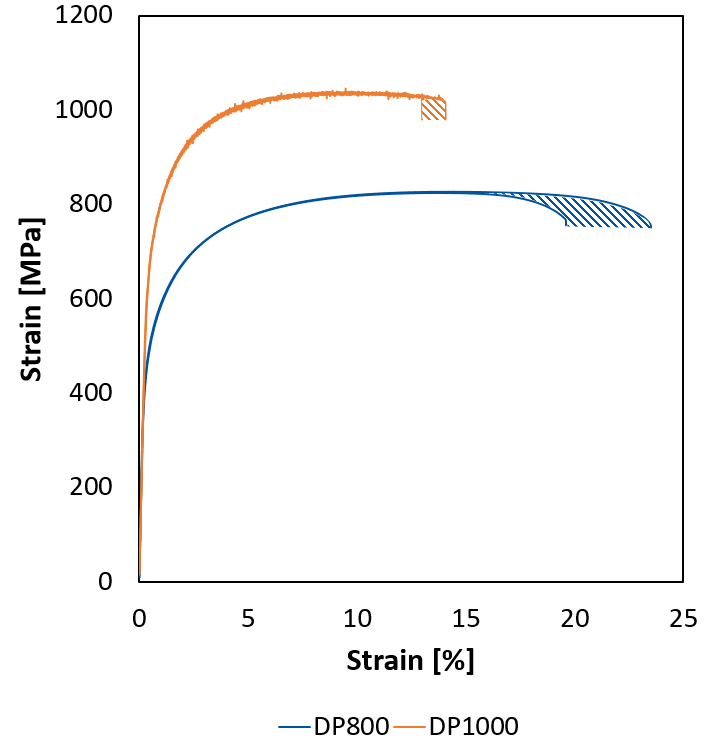}
\caption{Engineering stress-strain curve of uniaxial tensile tests of DP800 and DP1000}
\label{fig:2}       
\end{figure}

The results for both materials are shown in \textbf{Fig. \ref{fig:2}}. The scatter shown is the difference in fracture, resulting from the three tensile tests per material mentioned before. From this figure it is obvious, that DP1000 shows higher strength, while the strain at fracture of DP800 is considerably higher. The higher strength is a result of both; the higher martensite phase fraction as well as significantly refined grains in DP1000. \newline
In addition, there is a clear scatter in the elongation values, with elongation at fracture between 18 and 23 percent for DP800, while the scatter for the DP1000 is about one percent. This variation in elongation at fracture can be explained by the clearly pronounced band structures, which can lead to significant deviations depending on the position of the bands in the specimen. Since DP1000 fractures shortly after uniform strain, the necking is far less pronounced than in DP800. During necking the stress state in the sample can change quite significantly leading to non-proportional loading paths during the deformation of the sample. Thus, to describe the materials behaviour after the uniform strain, it is necessary for the material model to consider the effects of the changes of stress state during deformation. Therefore, a development of the existing MBW model was required to determine the material behaviour and damage accumulation more realistically.

\section{Extension of MBW model for non-proportional loading paths (npMBW-19)}
\label{MatModel}

In the framework of continuum damage mechanics, the modified Bai Wierzbicki (MBW) model has been proposed by Lian et al. (2013) and widely applied to describe the damage and fracture behaviours of various grades of steels \citep{Lian.2012, Munstermann.2017, Wu.2017, Novokshanov.2015, Shen.2020, Liu.2020}. Like in other damage mechanics models, the significant influences of stress state on the ductile fracture are considered through defining a strain based criterion which is usually a weighted function of two particularly important variables, the stress triaxiality $\eta$ and the Lode-angle parameter $\overline{\theta}$ that are related to the three stress invariants.

\begin{equation}
I_1 = tr[\sigma] = (\sigma_1 + \sigma_2 + \sigma_3)
\end{equation}

\begin{equation}
J_2 = \dfrac{1}{3}[\overline{\sigma}]^2 = \dfrac{1}{6} [(\sigma_1 - \sigma_2)^2 + (\sigma_2 - \sigma_3)^2 + (\sigma_3 - \sigma_1)^2]
\end{equation}

\begin{equation}
J_3 = \Big(\sigma_1 - \dfrac{I_1}{3}\Big)\cdot\Big(\sigma_2 - \dfrac{I_1}{3}\Big)\cdot \Big(\sigma_3 - \dfrac{I_1}{3}\Big)
\end{equation}

\begin{equation}
\begin{split}
\eta & = \dfrac{I_1}{\sqrt[3]{3 \cdot J_2}} = \dfrac{I_1}{3 \overline{\sigma}} \\
 & =  \dfrac{(\sigma_1 + \sigma_2 + \sigma_3)}{\sqrt[3]{\dfrac{1}{2} [(\sigma_1 - \sigma_2)^2 + (\sigma_2 - \sigma_3)^2 + (\sigma_3 - \sigma_1)^2]}}
\end{split}
\end{equation}

\begin{equation}
\theta = \dfrac{1}{3} cos^{-1} \Bigg ( \dfrac{3 \cdot \sqrt{3} \cdot J_3}{2 \cdot J_2^{3/2}} \Bigg )
\end{equation}

\begin{equation}
\overline{\theta} = 1 - \dfrac{6 \theta}{\pi}
\end{equation}

where $\sigma_1$,  $\sigma_2$ and $\sigma_3$ are principal stresses and $\overline{\sigma}$ is the von Mises equivalent stress. For the material model, the Lode-angle parameter $\overline{\theta}$ was used, which has a linear relationship with the Lode-angle $\theta$. \newline

The effects of stress state on plasticity in some metallic materials have been reported, while steels typically show a negligible pressure sensitivity, therefore, only the effects of Lode-angle parameter are considered in the yield criterion of the MBW model. 

\begin{equation}
\Phi = \overline{\sigma}(\sigma)- (1-D) \cdot \sigma_y ( \overline{\epsilon}^p,\overline{\theta}) \leq 0
\end{equation}

\begin{equation}
\sigma_y ( \overline{\epsilon}^p,\overline{\theta}) = \sigma_y ( \overline{\epsilon}^p) \cdot \Big [c_{\theta}^s + (c_{\theta}^{ax} - c_{\theta}^s) \cdot \Big ( \gamma - \dfrac{\gamma^{m+1}}{m+1} \Big) \Big]
\end{equation}

\begin{equation}
\gamma = \dfrac{\sqrt{3}}{2-\sqrt{3}} \cdot \Bigg [ sec \big( \dfrac{\overline{\theta} \cdot \pi}{6}\big)-1 \Bigg ]
\end{equation}

\begin{equation}
  c_{\theta}^{ax}=\begin{cases}
    c_{\theta}^{t}, & \text{for $\overline{\theta} \geq 0$}.\\
    c_{\theta}^{c}, & \text{for $\overline{\theta} < 0$}.
  \end{cases}
\end{equation}

Where $D$ is a scalar variable to quantify the damage effects, yield stress $\sigma_y$ is determined by the equivalent plastic strain $\overline{\epsilon}^p$ and Lode-angle parameter $\overline{\theta}$, and $\sigma_y ( \overline{\epsilon}^p)$ corresponds to the flow stress at given equivalent plastic strain $\overline{\epsilon}^p$ under the reference stress state.
$c_{\theta}^{s}, c_{\theta}^{t}, c_{\theta}^{c}$ are the normalised strength under shear, tension and compression state and $m$ is a material parameter with positive integral values that describes the Lode-angle sensitivity.
$\gamma$ is another stress state parameter with unique correlation to the Lode-angle parameter $\overline{\theta}$.
According to the derivation of Lian et al., \citep{Lian.2012}, the yield locus of MBW model is convex if the material parameters are located within the specific range of $\frac{\sqrt{3}}{2} \leq  \frac{c_{\theta}^{s}}{c_{\theta}^{ax}} \leq 1$. The conventional normality rule is applied in the MBW model and the plastic strain components are updated according to the following equation and $d\lambda$ is a non-negative plastic multiplier.

\begin{equation}
d\epsilon^p = d\lambda \cdot \dfrac{\delta \Phi}{\delta \sigma}
\end{equation}

In the coupled damage mechanics model, two individual criteria have been defined to identify the ductile damage initiation (DDI) and ductile fracture (DF), which corresponds to the initiation of degradation on microscopic scale in the material and the loss of load carrying capacity on the macroscopic scale. Numerically, damage initiation, in this study, is defined as the onset of macroscopic softening due to damage, which must be taken into account by the numerical representation of the material behaviour. In order to consider the change of stress state during plastic deformation, the average values of the stress triaxiality $\eta_{avg}$ and the Lode-angle parameter $\overline{\theta}_{avg}$ have been used to describe the stress state for non-proportional loading paths \citep{Wu.2017, Mu.2020}. 

\begin{equation}
\eta_{avg} = \dfrac{1}{\overline{\epsilon}^p} \int_0^{\overline{\epsilon}^p} \eta(\overline{\epsilon}^p) d \overline{\epsilon}^p
\end{equation}

\begin{equation}
\overline{\theta}_{avg} = \dfrac{1}{\overline{\epsilon}^p} \int_0^{\overline{\epsilon}^p} \overline{\theta}(\overline{\epsilon}^p) d \overline{\epsilon}^p
\end{equation}
\newline
Since the damage is dependent on stress state it is necessary to define equations for the initiation of damage, as well as the fracture, that represent this dependency. Therefore, the damage initiation locus (DIL) and ductile fracture locus (DFL) have been defined as two individual equations $f_{di}$ and $f_{df}$ with the stress triaxiality and the Lode-angle parameter as independent variables. The instantaneous and average values of the independent stress state variables have been used in the damage and fracture criteria under proportional and non-proportional loading conditions, respectively. Under non-proportional loading conditions, these two equations describe the critical equivalent plastic strains at the moment of damage initiation and ductile fracture, respectively.



\begin{multline}
f_{di}(\eta_{avg},\overline{\theta}_{avg}) = \Big[\dfrac{1}{2}(D_1 e^{-D_2 \eta_{avg}} + D_5 e^{-D_6 \eta_{avg}}) \\
- D_3 e^{-D_4 \eta_{avg}} \Big] \overline{\theta}_{avg}^2 + \dfrac{1}{2} (D_1 e^{-D_2 \eta_{avg}} - D_5 e^{-D_6 \eta_{avg}}) \overline{\theta}_{avg} \\
+ D_3 e^{-D_4 \eta_{avg}}
\end{multline}

\begin{multline}
f_{df}(\eta_{avg},\overline{\theta}_{avg}) = \Big[\dfrac{1}{2}(F_1 e^{-F_2 \eta_{avg}} + F_5 e^{-F_6 \eta_{avg}}) \\
- F_3 e^{-F_4 \eta_{avg}} \Big] \overline{\theta}_{avg}^2 + \dfrac{1}{2} (F_1 e^{-F_2 \eta_{avg}} - F_5 e^{-F_6 \eta_{avg}}) \overline{\theta}_{avg} \\
+ F_3 e^{-F_4 \eta_{avg}}
\end{multline}

where $D_1 $ - $ D_6$ and $F_1 $ - $ F_6$ are material parameters used to define the damage initiation locus  and ductile fracture locus. Under the condition that $D_1 = D_5,D_2=D_6$ and $F_1=F_5,F_2=F_6$, the DIL and DFL are symmetric with respect to the Lode-angle parameter and four independent parameters are enough to define the corresponding loci. Based on previous experimental observations, a cut-off value of the stress triaxiality $\eta_c$ exists, below which the initiation and evolution of ductile damage cannot be triggered due to pressure effects. $\eta_c = \frac{-1}{3}$ as a reasonable estimation has been adopted in the MBW model \citep{Wu.2017}. Therefore, when the stress triaxiality is lower than $\eta_c$, the equations $f_{di}$ and $f_{df}$ are set to be infinite. The damage initiation specified by this model is unrelated to the materials mechanisms of damage initiation, e.g. micro crack formation, void formation. Instead, it aims to describe the aggregative accumulation of the defects and their influence on the load bearing capabilities. For this step a plasticity model is no longer able to describe the materials mechanical behaviour \citep{Keim.2019}. For the non-proportional loading, two indicators have been applied to describe the ductile damage initiation $I_{dd}$ and ductile fracture $I_{df}$ respectively to consider the effects of stress state evolution.

\begin{multline}
I_{dd} = \int_0^{\overline{\epsilon}^p} \dfrac{d\overline{\epsilon}^p}{\overline{\epsilon}_{di}^{p}(\eta_{avg},\overline{\theta}_{avg})} ~~~~~~ \text with\\
\overline{\epsilon}^{p}_{di}(\eta_{avg},\overline{\theta}_{avg})=\begin{cases}
    + \infty , & \eta_{avg} \leq \eta_c \\
    f_{di}(\eta_{avg},\overline{\theta}_{avg}), & \eta_{avg} > \eta_c.
    \end{cases}
\end{multline}

\begin{multline}
I_{df} = \int_{\overline{\epsilon}^{p,c}_{di}}^{\overline{\epsilon}^p} \dfrac{d\overline{\epsilon}^p}{\overline{\epsilon}^{p}_{df}(\eta_{avg},\overline{\theta}_{avg})} ~~~~~~ \text with\\
\overline{\epsilon}^{p}_{df}(\eta_{avg},\overline{\theta}_{avg})=\begin{cases}
    + \infty , & \eta_{avg} \leq \eta_c \\
    f_{df}(\eta_{avg},\overline{\theta}_{avg}), & \eta_{avg} > \eta_c.
    \end{cases}
\end{multline}

The values of equivalent plastic strain and equivalent stress at the moment of damage initiation ($I_{dd} = 1$) are defined as two characteristic variables $\overline{\epsilon}_{di}^{p,c}$ and $\overline{\sigma}^{c}_{di}$, respectively:

\begin{equation}
\overline{\epsilon}_{di}^{p,c} = \overline{\epsilon}^{p} ~~~~~~ (I_{dd} = 1)
\end{equation}

\begin{equation}
\overline{\sigma}_{di}^{c} = \overline{\sigma} ~~~~~~ (I_{dd} = 1)
\end{equation}

After the damage initiation criterion is fulfilled, damage evolution is controlled according to the energy dissipation theory. Depending on the shape of damage initiation locus and ductile fracture locus, when the indicator of the ductile fracture $I_{df}$ reaches unity, the damage variable $D$ does not necessarily reach unity. Therefore, a critical value of the damage variable $D_{cr}$ exists, at which the material point will fail regardless of the value of the $D$ variable:

\begin{equation}
D_{cr} = \dfrac{\overline{\sigma}_{di}^{c}}{G_f}\big(\overline{\epsilon}_{df}^p - \overline{\epsilon}_{di}^p \big )
\end{equation}

Where $G_f$ is a material parameter which controls the damage evolution rate. Linear damage evolution is assumed in the MBW model, which is expressed as:

\begin{equation}
  D=\begin{cases}
    0, & I_{dd} < 1\\
    D_{cr} \cdot I_{df}, & I_{dd} \geq 1 ~~~\wedge~~~ I_{df} < 1 \\
    1, & I_{dd} \geq 1 ~~~\wedge~~~ I_{df} \geq 1 
  \end{cases}
\end{equation}

In summary, the damage evolution is determined by the two independent damage initiation and ductile fracture criteria. After a certain damage nucleation period, which is controlled by plastic deformation, damage evolution takes place. Once the indicator of the ductile fracture $I_{df}$ reaches unity, the final crack propagation is triggered and failure occurs. Therefore, the model, hereafter called npMBW-19, is capable of representing the influence of the necking, and thus the change of stress state, during deformation.

\section{Calibration of the new model for materials DP800 and DP1000}
\label{MatModelCalib}

\begin{figure*}
   \includegraphics[width=\textwidth]{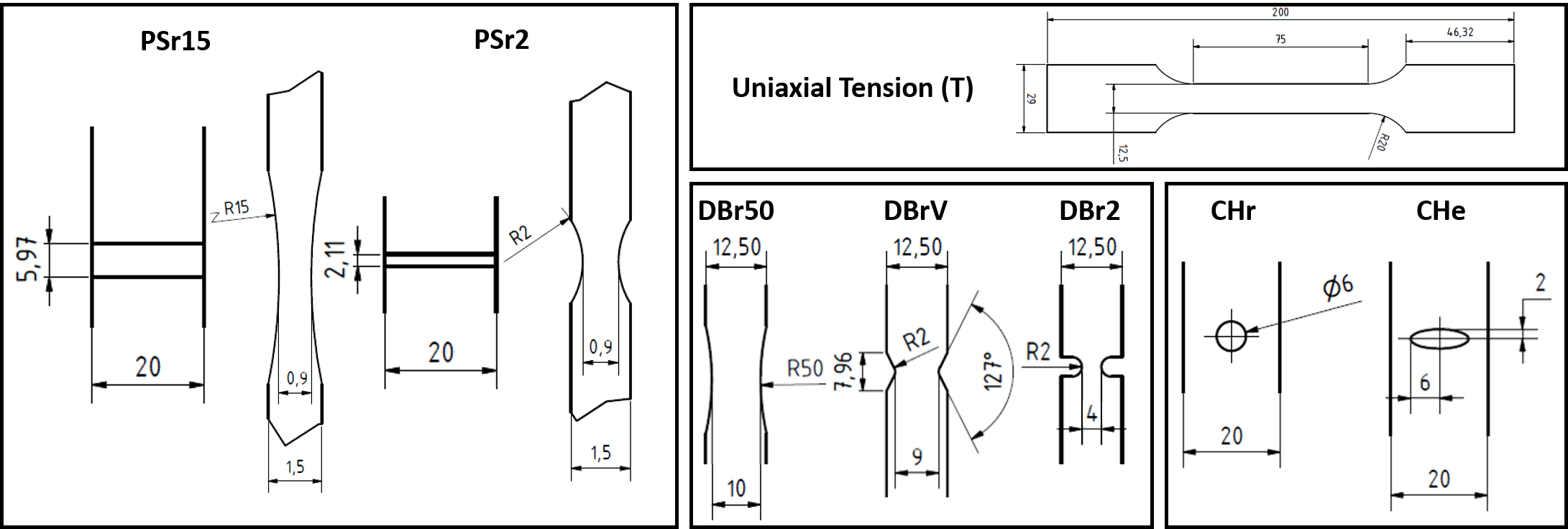}
\caption{Applied tensile specimens for steel DP800}
\label{fig:11}       
\end{figure*}

\begin{figure*}
   \includegraphics[width=\textwidth]{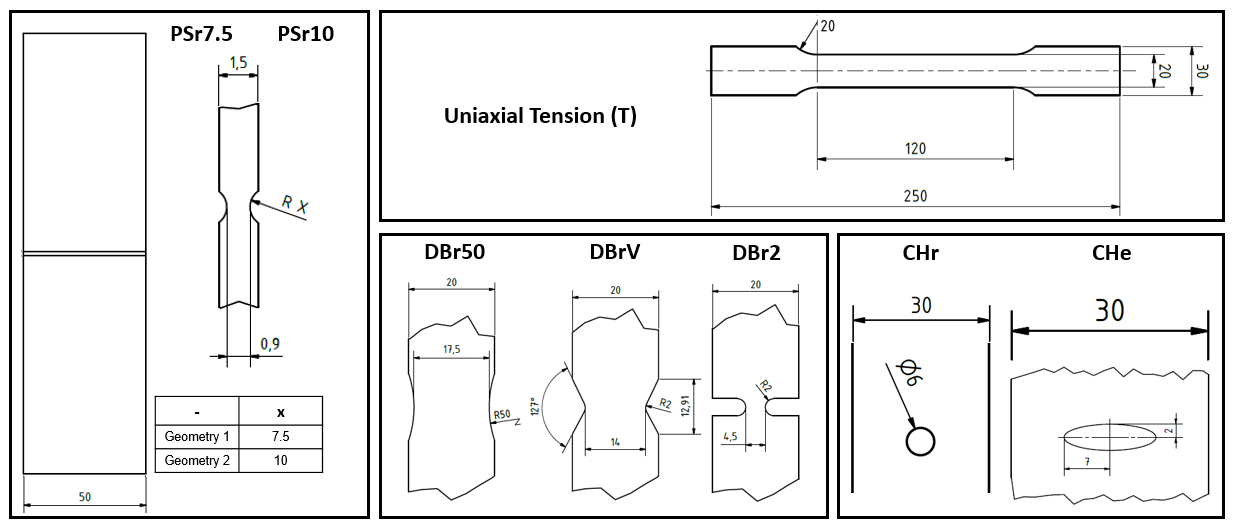}
\caption{Applied tensile specimens for steel DP1000}
\label{fig:12}       
\end{figure*}

The calibration approach for the material models for both steels follows roughly the approach of Lian et al. \citep{Lian.2012, Lian.2014}. Since the calibrated npMBW-19 model needs to be able to account for various stress states, the calibration of the material model is carried out on a variety of sample geometries. By varying the sample geometries in tensile tests, different stress states can be accomplished. In this study three differently notched specimen types were applied for the calibration of the material model in addition to the uniaxial tensile test. Used specimen types were: Notched dogbone samples (varying notch geometries at the edge of the sample), central hole samples (round, as well as elliptical holes in the center of the specimen) and plane strain samples (notch with different radii over the thickness of the sample). The applied specimens for each material can be seen in \textbf{Fig. \ref{fig:11}} and \textbf{ Fig. \ref{fig:12}} The type of notch of the sample is abbreviated with an r continuing with the radius, for the notched dog bone samples.

The corresponding stress states, characterised by the Lode-angle parameter and the stress triaxiality in the applied samples are delineated in \textbf{Table \ref{tab:2}}. To achieve multiple stress states, notches were modified with various radii to gain geometries of different stress states within one sample type. Per specimen geometry, three tensile tests were performed in accordance with the procedure described earlier for the uniaxial tensile test. Afterwards, simulations of the experiments were conducted,  using ABAQUS, to achieve a comparison between the force - displacement curves of experimentally determined values and simulated ones.

\begin{table}
\caption{Stress states of utilised sample geometries}
\label{tab:2}       
\begin{tabular}{lll}
\hline\noalign{\smallskip}
\textbf{Sample geometry}& \textbf{Lode-angle} & \textbf{Stress triaxiality $\eta$}  \\
& \textbf{parameter $\overline{\theta}$} & \\
\noalign{\smallskip}\hline\noalign{\smallskip}
Uniaxial tensile (UT) & 1 & $\frac{1}{3}$  \\
Notched dog bone (DB) & 0.3 - 0.8 & 0.4 - 0.6  \\
Central hole (CH) & \textasciitilde 1 & 0.3 - 0.4  \\
Plane strain (PS) & 0 & 0.5 - 0.7  \\
\noalign{\smallskip}\hline
\end{tabular}
\end{table}

For the determination of the base flow curve, the uniaxial tensile tests (T), presented in the previous chapter, were utilised. From the determined engineering stress-strain curve, the true stress-true strain curve was calculated until the uniform elongation point. This data was then used to fit the Hollomon - Voce hardening model to the material's flow curve via the Matlab curve fitting tool.

\begin{equation}
\overline{\sigma} = \alpha \cdot (K \overline{\epsilon}_p^n) + (1- \alpha) \cdot (A - B \cdot e^{-C \overline{\epsilon}_p})
\end{equation}

This specific hardening model was chosen, since it shows a good compromise between accurate representation at low plastic strains and realistic hardening behaviour for higher strains. In \textbf{Table \ref{tab:3}} the parameters for the Hollomon-Voce models are given for both, DP800 and DP1000.

\begin{table}
\caption{Hollomon-Voce fitting parameters for steels DP800 and DP1000}
\label{tab:3}       
\begin{tabular}{llllllll}
\hline\noalign{\smallskip}
& $\alpha$ & $K$ & $n$ & $A$ & $B$ & $C$     \\
\noalign{\smallskip}\hline\noalign{\smallskip}
DP800 & 0.5138 & 1843 & 0.44 & 1167 & 820.4 & 100   \\
DP1000 & 0.5879 & 2000 & 0.1127 & 725.5 & 300 & 57.2   \\
\noalign{\smallskip}\hline
\end{tabular}
\end{table}

After the calibration of the flow curve the basic parameters of the MBW model were determined ($c_{\theta}^s$, $c_{\theta}^t$, $c_{\theta}^c$, m). This was done by iterating over multiple simulations using a range of different sample geometries.\newline

Subsequently the damage and fracture parameters of the npMBW-19 model were determined. Damage and fracture criteria in this material model are described by equations 14 and 15. Therefore, the specified locus needs to be calibrated for both events, damage initiation and fracture \citep{Lian.2012, Wu.2017}.  For the damage initiation locus, a comparison of force and displacement curve between simulation and experimental results was used. Since the damage described in this model is related to the accumulated damage incidents, a threshold method has been utilised to find the numerical damage initiation. For that reason, the numerical onset of damage was determined as the point where the deviation between simulated and experimental force and displacement curves was apparent. Similar methods have been used by other authors within the damage mechanics field \citep{Brvik.2001, Bouchard.2011}. \newline

At this step, the Lode-angle parameter and stress triaxiality as well as the equivalent plastic strain (PEEQ) are taken from the simulation. Since these parameters may vary locally, the element is chosen that shows the most critical state of stress and thus is most likely to encounter damage first. By extracting the Lode-angle parameter, stress triaxiality and equivalent plastic strain for a multitude of different tensile geometries, data points are gathered in the space defined by these three variables. Applying the curve fitting tool of \textit{Matlab}, a function can be defined that describes the desired surface while using the obtained results as supporting points. For non-proportional loading paths it is necessary to average the stress state of the critical element, where damage happens first, over the simulated steps \citep{Wu.2017}.

After determining the locus for the onset of damage (DIL), the effect that damage has on the component needs to be adjusted.
In the npMBW-19 model parameter $G_f$ is calibrated to adjust the speed at which damage accumulates in the simulated material.
$G_f$ is defined as the energy dissipation between damage initiation and complete fracture.
When the softening is specified the fracture locus can be determined. The approach used for this determination follows the one from the damage initiation locus closely. This time the point for the experimental fracture is compared to the simulation. The step where the fracture should occur is identified and Lode-angle parameter, stress triaxiality and equivalent plastic strain are extracted for the critical elements. Again, the stress states are averaged from the point of damage initiation to the presumed fracture of the sample. After gathering the data for all sample geometries the locus is fitted in regards to the obtained points using the \textit{Matlab} curve fitting tool.

\begin{figure} [htbp]
   \includegraphics[width=0.5\textwidth]{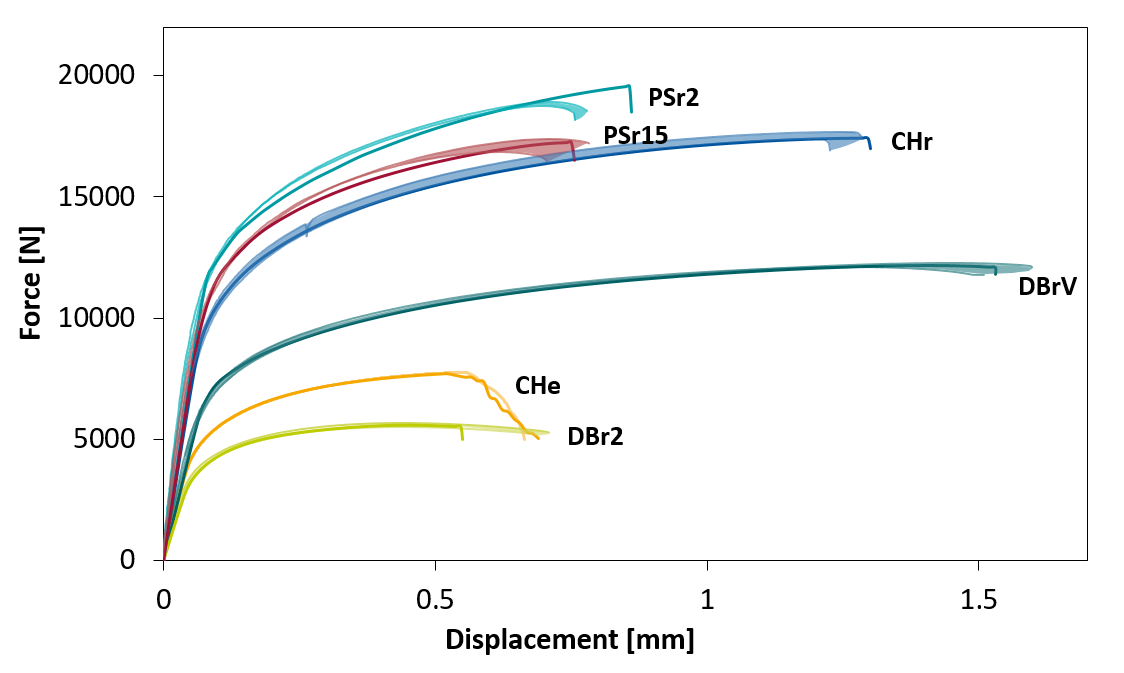}
\caption{Comparison between experimentally obtained data (background and lighter color) and simulation results for DP800}
\label{fig:3}       
\end{figure}

In \textbf{Fig. \ref{fig:3}} the final results of this calibration process are depicted for steel DP800. From this figure it is obvious, that a good match between experimental data and simulations was been obtained. The scatter for the experimental testing can be seen in the shaded areas.

\begin{figure}[htbp]
   \includegraphics[width=0.5\textwidth]{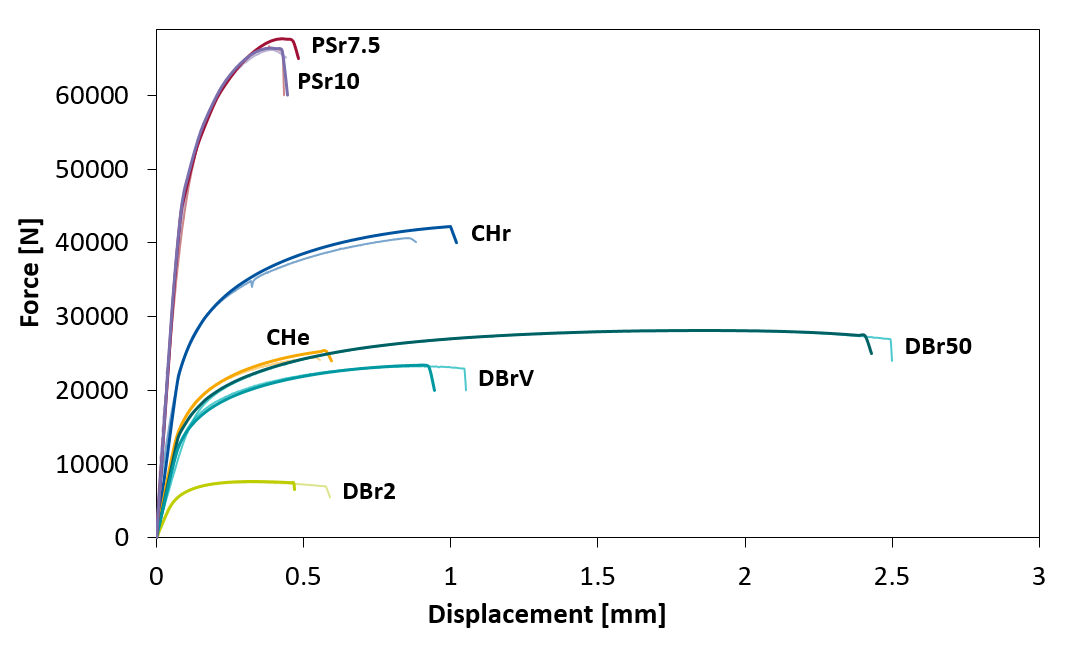}
\caption{Comparison between experimentally obtained data(background and lighter color) and simulation results for DP1000}
\label{fig:4}       
\end{figure}

Likewise, the material model for DP1000 was calibrated. The applied flow curve can be seen in \textbf{Table \ref{tab:3}}. Additionally, a damage initiation, as well as a ductile fracture locus were calibrated using the same approach as described above for the DP800. By duplicating the approach stated above, a good agreement with the experimental data could be reached (\textbf{Fig. \ref{fig:4}}). Contrary to the DP800 almost no scatter could be found during the tests of the DP1000 material, which also shows no significant banding in its microstructure. The applied set of parameters can be found in \textbf{Table \ref{tab:4}}. Interestingly, the calibrated $G_f$ parameter for DP1000 is higher which results in a slower development of the damage variable. This results in a fairly slow accumulation of damage after the initiation.

\begin{table}
\caption{Calibrated npMBW-19 parameter sets for both materials}
\label{tab:4}   
\begin{tabular}{llllllll}
\hline\noalign{\smallskip}
& $c_{\theta}^s$ & $c_{\theta}^t$ & $c_{\theta}^c$ & m & $D_1$ & $D_2$ & $D_3$    \\
\noalign{\smallskip}\hline\noalign{\smallskip}
DP800 & 0.95 & 1 & 0.9 & 6 & 0.5 & 2 & 0.365  \\
DP1000 & 0.95 & 1 & 0.97 & 6 & 0.4 & 1 & 0.1  \\
\noalign{\smallskip}\hline
& $D_4$ & $G_f \big[\frac{J}{mm^3}\big]$ & $F_1$ & $F_2$ & $F_3$ & $F_4$ &     \\
\noalign{\smallskip}\hline\noalign{\smallskip}
DP800 & 3 & 1.2 & 0.7 & 1 & 0.366 & 2 &  \\
DP1000 & 1.5 & 6.5 & 0.58 & 0.76 & 0.443 & 1.57 &  \\
\end{tabular}
\end{table}

\section{Damage behaviour prediciton}
\label{DamBeha}

For the scope of this study, it is important to differentiate between failure and damage of a material or component. Because damage is the deterioration of materials properties before failure, especially the load bearing capacity \citep{Lemaitre.1992}, it is not to be equated with component failure. Damage occurs on a microscale and is usually described as the development of voids inside the microstructure, while on a macroscale damage usually equates to cracks in the component and therefore can be seen as component failure. It is therefore highly relevant to differentiate between micro and macroscale \citep{Tekkaya.2017}. For numerical analysis, damage is defined as the macroscopic reduction of the stress during loading, that cannot be described by basic plasticity modelling. Thus, Lemaitre introduced a factor for damage in a microstructure, which results in a reduction of the flow potential by the term ($1-D$), where $D$ is the damage variable \citep{Lemaitre.1985}. The damage variable adopted by Lian et al. shows some differences to the one postulated by Lemaitre. While Lemaitre’s damage variable is calculated based on the area fraction of defects, Lian et al. refer to the stress at damage initiation, divided by the energy required to create new surfaces in a volume of the material (see $G_f$), an adaptation of the damage evolution law used by ABAQUS finite element code \citep{Lian.2012}. Accordingly, both damage variables are scalar, but there are quite pronounced differences between both numerical damage rules. These differences between the damage models must be distinguished, as well as the differences between micro- and macroscopic damage phenomena. \newline

\begin{figure*}[htbp]
   \includegraphics[width=\textwidth]{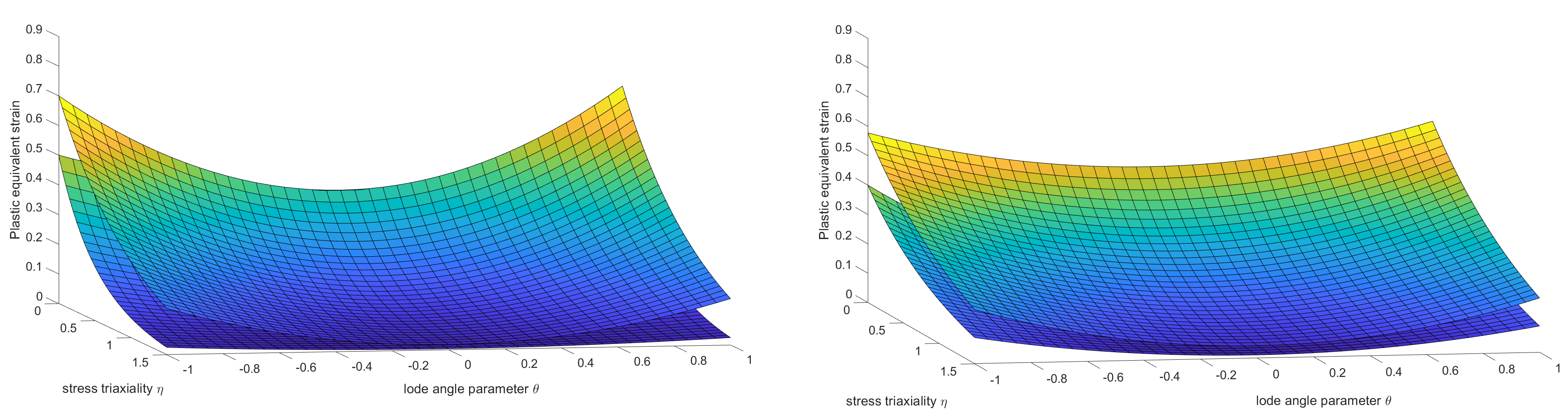}
\caption{Comparison of the ductile damage initiation locus and the ductile fracture locus for DP800 (left) and DP1000 (right)}
\label{fig:5}       
\end{figure*}

Because the damage law used in the MBW model does not refer to a physical material characteristic, like the area fraction of voids, except for the energy for cracks, it is a rather more phenomenological approach to model the influence of damage on the materials flow potential. The damage evolution, as discussed before, starts when a specific equivalent plastic strain (PEEQ) locally exceeds a certain threshold, which changes with stress states. The respective value for PEEQ is determined by the damage initiation locus. After this point, softening occurs in the simulation, which leads to a direct reduction of the resulting stress compared to virgin materials. The length of this following phase where damage accumulates depends on the stress state which is considered in the ductile fracture locus. The comparisons of the ductile damage initiation locus and the ductile fracture locus for each respective material are shown in \textbf{Fig. \ref{fig:5}}. The shape of the loci for DP800 and DP1000 are different, as was to be expected. The distance between the plots is higher for the DP 800 material which leads to a longer damage accumulation phase. Merely for higher triaxialities and Lode-angle parameters around zero, the differences between the loci of DP800 and DP1000 is minimal. Some research suggests a different shape for the ductile damage initiation locus and the ductile fracture loci, especially for the area around a stress triaxiality of 0, namely shear stress state \citep{Papasidero.2015}. Nevertheless, based on the experimental and numerical results in this study, the loci in \textbf{Fig. \ref{fig:5}} constructed for both steels using the corresponding calibrated damage and fracture parameters is validated within the range of investigated stress states. In the case of an application of the calibrated material model for even lower or higher stress triaxialities, the loci would have to be revisited to confirm or adapt their shape.

\begin{figure}[htbp]
   \includegraphics[width=0.5\textwidth]{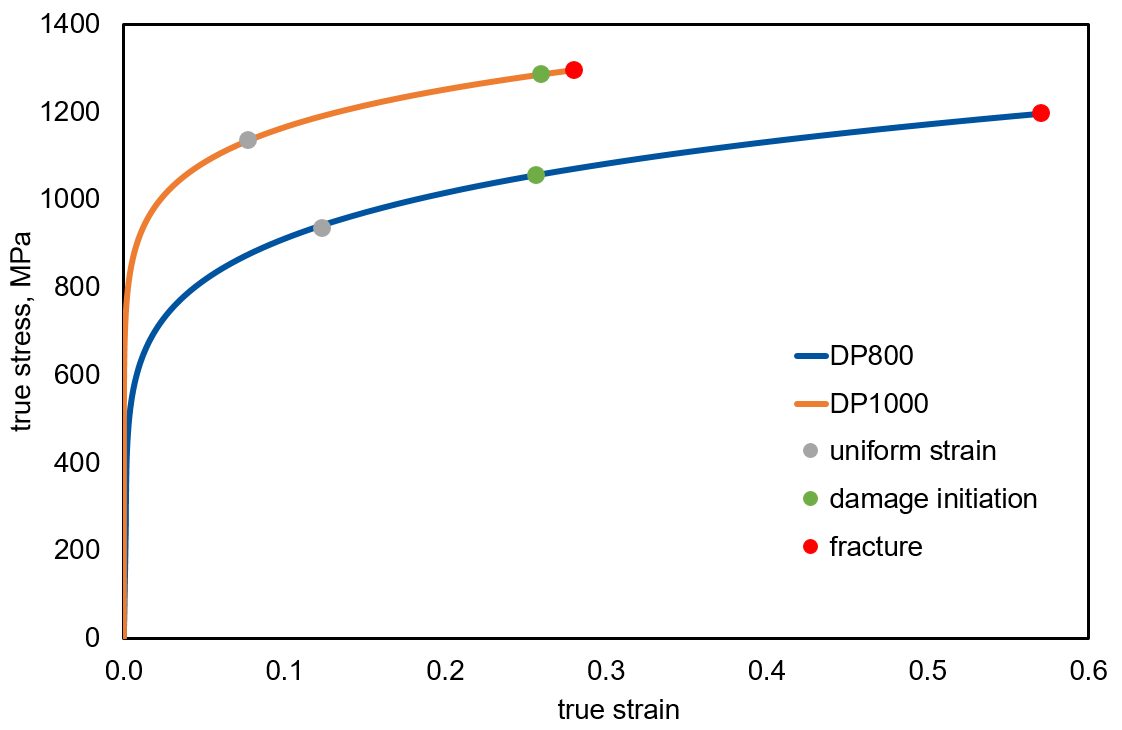}
\caption{Flow curves from uniaxial tensile test of DP800 and DP1000 with numerically determined points of damage initiation and fracture under uniaxial tension condition.}
\label{fig:6}       
\end{figure}

Due to the differences in the distance of the DIL and the DFL the damage accumulation phase is significantly different between the two steels. \textbf{Fig. \ref{fig:6}} shows this difference utilising the flow curves obtained from the uniaxial tensile test of both materials, as well as calculating the points for damage initiation and fracture under uniaxial tension condition $(\eta = \frac{1}{3} , \overline{\theta} = 1)$ based on calibrated material parameters. The point for the damage initiation takes place at roughly the same strain for both materials, while fracture is delayed significantly for DP800.

To further examine precision of the numerical results, interrupted tensile tests were conducted for both materials. For each material a sample was therefore first tested until failure and subsequent specimens of identical geometry were stopped after a distinct strain was reached. The lowest elongation used in this investigation was the uniform elongation, as no or little damage is expected below this. This way a metallographic damage analysis could be carried out to investigate the average amount of damage that could be observed in a sample. For both materials unnotched dog bone specimens were utilised to ensure a good comparability. For the analysis of the damage in the material, light optical microscopy was chosen, since a bigger area can be investigated by light optical analysis, where scanning electro microscopy (SEM) pictures resolve only smaller areas of the samples. However, it is not easily possible to differentiate between voids and inclusions in the material. Thus the area fraction that is detected is not quantitatively representative of the actual void fraction. To find out about the area fraction for each sample, multiple pictures were taken to gather information about the scatter band where the actual values lie. For this analysis, the light optical pictures were converted to greyscale images, which were subsequently evaluated by a threshold method, with which a differentiation between matrix material and voids/inclusion could be made. For these steps \textit{Fiji} was used as image analysis software \citep{Rueden.2017, Schindelin.2012}.

\begin{figure*}[htbp]
   \includegraphics[width=\textwidth]{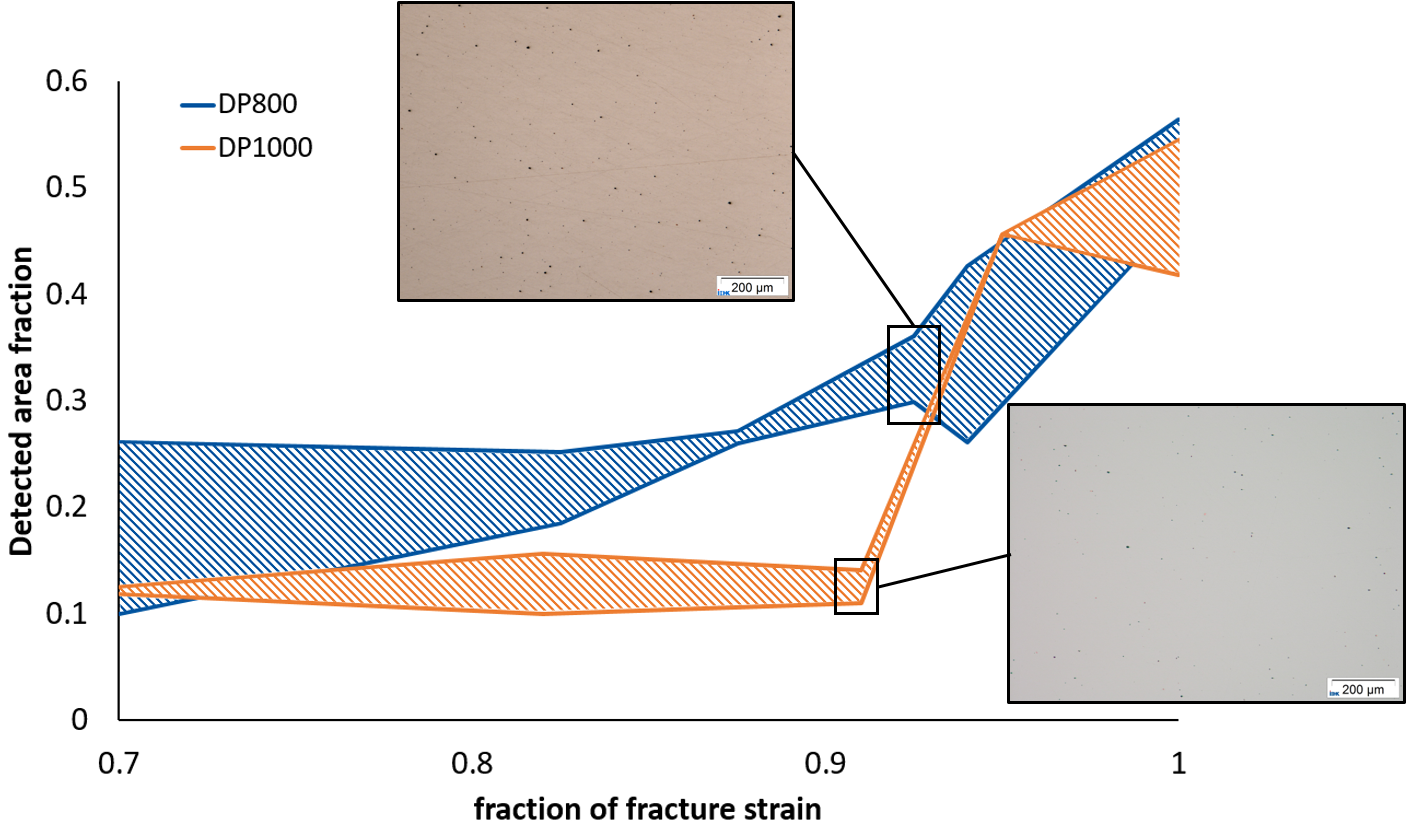}
\caption{Comparison of light optically detected area fraction for DP800 and DP1000 for different strains}
\label{fig:7}       
\end{figure*}

The results of this analysis are depicted in \textbf{Fig. \ref{fig:7}}. To better compare the values for both materials, a normalisation was carried out, where the current strain was divided by the respective fracture strain. A comparison of the values for the detected area fraction reveals a gradual increase for DP800, while for DP1000 no significant rise in fraction can be observed until just before fracture of the sample. The large scatter, especially at the beginning, can be explained by the lack of necking, which means that the region of interest cannot be identified accurately.

\begin{figure}[htbp]
   \includegraphics[width=0.5\textwidth]{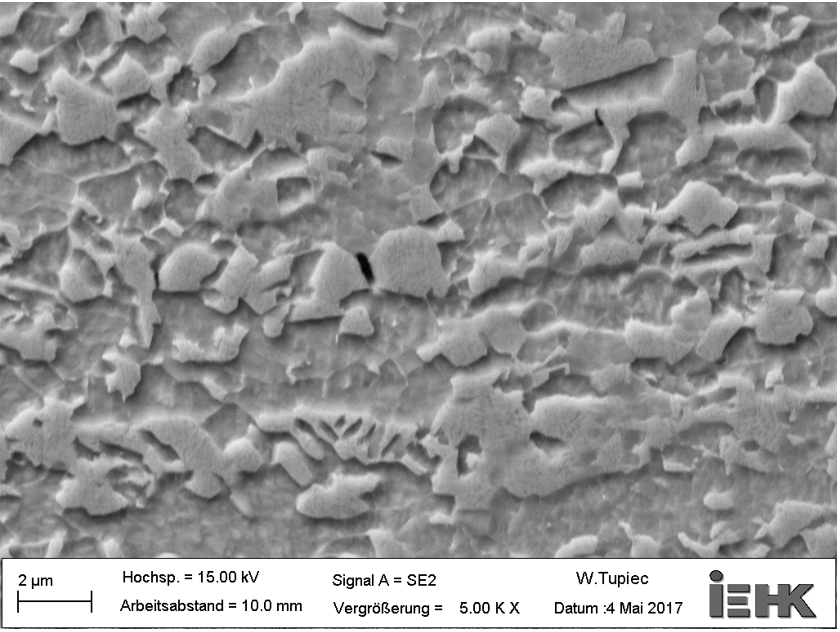}
\caption{Evolution of damage in DP1000. Single martensite crack with magnification of 5000 for 0.7 of fracture strain}
\label{fig:8}       
\end{figure}

\begin{figure}[htbp]
   \includegraphics[width=0.5\textwidth]{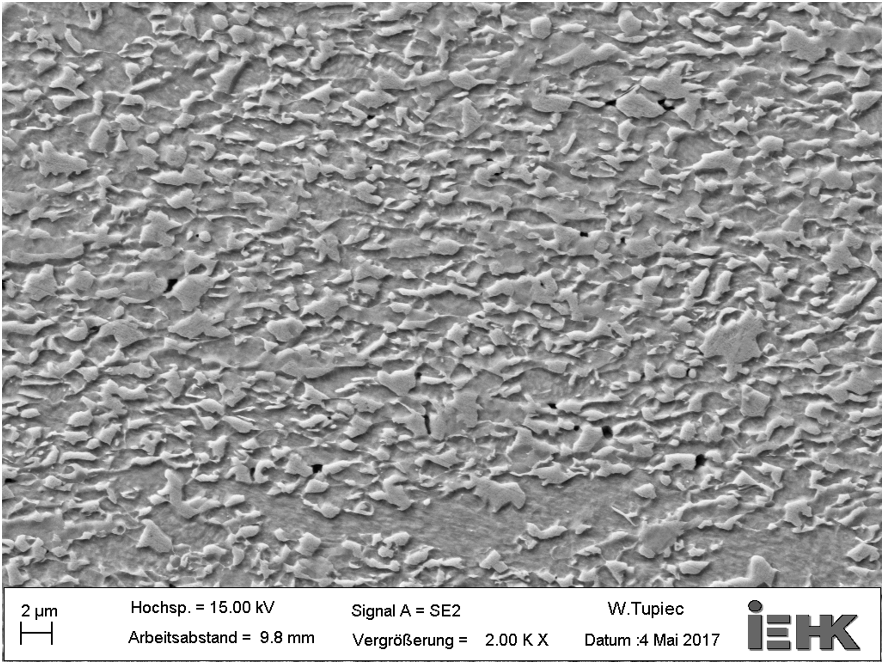}
\caption{Evolution of damage in DP1000. Many voids have formed and grown, magnification of 2000, 0.95 of fracture strain}
\label{fig:9}       
\end{figure}

Thus, the damage accumulation phase for DP800 starts at lower strains relative to the fracture strain of the material. By contrast, the damage accumulation phase for DP1000 starts very late and just before fracture. Therefore, the damage in the material behaves exactly as predicted using the npMBW-19 model.
To assess the damage state in the microstructure, pictures were taken in the SEM. Especially for DP1000 an analysis for higher magnifications was necessary to reveal if damage forms earlier than shortly before fracture. For 70\% of the fracture strain, only very few events of damage initiation could be found under high magnification (\textbf{Fig. \ref{fig:8}}). While the amount of these initiation locations increases with the strain, growth is very limited (\textbf{Fig. \ref{fig:9}}).

\begin{figure}[htbp]
   \includegraphics[width=0.5\textwidth]{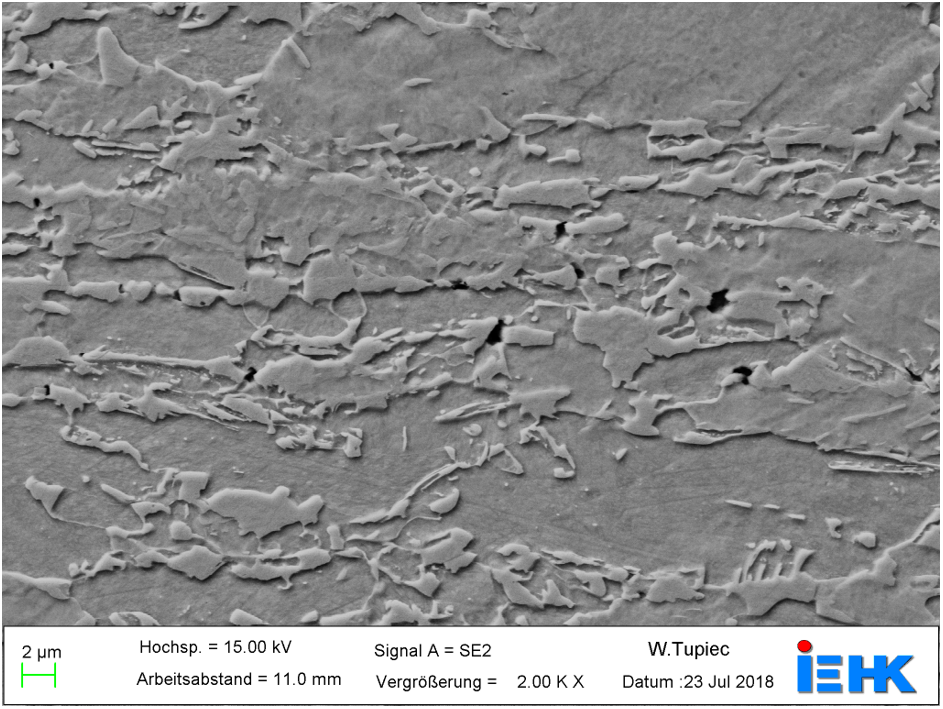}
\caption{Damage shortly before fracture in DP800}
\label{fig:10}       
\end{figure}

A comparison of the samples that are at 95\% of fracture strain reveals, that the voids in DP1000 (\textbf{Fig. \ref{fig:9}}) are significantly smaller than in DP800 (\textbf{Fig. \ref{fig:10}}). It is therefore concluded, that the damage accumulation phase for DP1000 is indeed significantly shortened. In particular, it is noticeable, that no void in DP1000 exceeds a length of 1$\mu$m , while the DP800 features multiple larger voids. Additionally, voids for DP800 are more circular, while they are shaped like cracks for DP1000 again leading to the conclusion, that there has been no time for growth after initiation. This is in line with the results demonstrated in \textbf{Fig. \ref{fig:6}}, where a shorter damage accumulation phase is present in DP1000 and thus a lower decrease of load bearing capabilities is to be expected.

Besides, the damage initiation modes were investigated. For both materials, the prevalent modus for damage initiation was the cracking of martensite islands. For DP800 the martensite bands especially were sites for damage initiation. Furthermore, decohesion of ferrite and martensite islands was found in the DP800 after about 80\% of fracture strain. 

\section{Conclusions}
\label{Conc}

This study showed significant differences between two industrially produced dual-phase steels. Starting with the experimental results, the difference in fracture strain was found to be significant with large scatter for the DP800. This scatter was attributed to the pronounced banding found within the material. For the numerical analysis it was found, that the change in the stress state during necking needs to be considered by the material model for proper simulation results. Therefore, the MBW model was extended to account for non-proportional loading paths. The parameters of the model were then fitted for both materials to reveal the disparities in the material behaviour numerically. Subsequently, the damage initiation and fracture loci were calibrated. The comparison of simulation results to the experimentally obtained force-displacement curves reveals a high agreement for both materials. Especially the differences in damage behaviour were modelled precisely. \newline

The found differences during the experimental testing and analysis can be attributed to the differences in the microstructure. Especially grain size and martensite content, but also the pronounced banding in the DP800 play an important role for mechanical properties, as well as damage initiation and accumulation. In this study, it was shown, that the damage in both dual-phase steels initiates at similar equivalent plastic strains. Oppositely, the fracture happens at vastly different equivalent plastic, as well as global strains. This leads to completely different damage accumulation phases in the material. The numerical simulations showed an exceedingly different length of the damage accumulation phase for the two steels. This difference was subsequently verified by experimental tests, where the amount of damage in the material after an interrupted tensile test was examined. For these tests it could be shown, that DP800 exhibits a pronounced damage accumulation phase, while DP1000 fractures shortly after a critical amount of voids forms in the material. Thus fracture occurs with almost no damage accumulation and the void growth phase is nearly skipped. Therefore, the different microstructures lead to specific damage characteristics, which in turn influence and change the specific properties of the material. Additionally, the contrast in the mechanical properties between the two phases for both materials reinforce this effect. Since the carbon content in martensite is relatively higher in DP800 than in DP1000, the martensite fractures earlier leading to a relatively early damage initiation and longer damage accumulation phase. For the DP1000 the the contrasts are not so distinct, which leads to a comparatively late initiation of damage and failure shortly afterwards.   \newline

During the comparison of experimental and numerical results, it was quite obvious, that the presented material model is able to accurately represent the experimental tensile test curves, both uniaxial and notched specimens. The stress states do, however show only minor variance. Thus, for higher deviations, an adjustment of the averaging scheme for determining stress triaxiality and Lode angle parameter for the ductile damage and fracture loci might be appropriate. Furthermore, the model shows a discontinuity around the value of $\eta_c = \frac{-1}{3}$, since a fixed value at which no damage is applied in the model, will be difficult to deal with, when getting close to it (e.g. $\eta = -0.32$). However, since the loading paths in this study are exclusively above this value, this is a promising and important concern for future development.

The comparison of numerical and experimental ductile damage showed, that the presented material model is able to accurately predict the damage initiation, damage accumulation and fracture of both materials. Nevertheless, the accuracy of the damage initiation point in the material model is still an important topic for further investigation. Since the location of the fracture locus strongly depends on the damage initiation locus, a high precision for the DIL is desirable. However, the commonly used method of direct current potential drop (DCPD) is not feasible for DP steel, since its void volume is comparably low. Therefore, an improvement of the method to determine aforementioned damage initiation locus is necessary and currently examined. One possibility is to measure the density of the material to determine the time of damage initiation \citep{Hering.2019,Schowtjak.2019, Meya.2019}. \newline

The analysis of damage initiation point and damage accumulation by light optical microscopy is rather qualitative than quantitative. Since statistical representativeness and accuracy have to be balanced for this type of examination, micro voids are not detected in the pictures. Thus, the values received are not the void area fraction. On top of that, light optical pictures show inclusions in a similar colour to voids, isolation of voids for analysis purpose is rather difficult for light optical microscopy. For a more quantitative result of void area fraction SEM pictures with high resolution over a big area seem to be more promising. \newline

The presented results suggest that the damage, and therefore the materials mechanical properties depend on the microstructure of the respective steel. It is therefore of high interest to investigate the influence of each microstructural parameter on the damage characteristics as well as the mechanical properties of the material. Three-dimensional representative volume elements seem to be a promising approach to investigate the influence of different microstructural characteristics, like martensite volume content, martensite morphology and grain size. 

\section{Acknowledgement}

The funding of this research by the Deutsche Forschungsgemeinschaft (DFG, German Research Foundation) – Projektnummer 278868966 – TRR 188, is gratefully acknowledged.

\bibliographystyle{spbasic}      
\bibliography{Bibliography2}   

\begin{thebibliography}{45}
\providecommand{\natexlab}[1]{#1}
\providecommand{\url}[1]{{#1}}
\providecommand{\urlprefix}{URL }
\expandafter\ifx\csname urlstyle\endcsname\relax
  \providecommand{\doi}[1]{DOI~\discretionary{}{}{}#1}\else
  \providecommand{\doi}{DOI~\discretionary{}{}{}\begingroup
  \urlstyle{rm}\Url}\fi
\providecommand{\eprint}[2][]{\url{#2}}

\bibitem[{Ahmad et~al.(2000)Ahmad, Manzoor, Ali, and Akhter}]{Ahmad.2000}
Ahmad E, Manzoor T, Ali KL, Akhter JI (2000) Effect of microvoid formation on
  the tensile properties of dual-phase steel. Journal of Materials Engineering
  and Performance 9(3):306--310

\bibitem[{Avramovic-Cingara et~al.(2009)Avramovic-Cingara, Ososkov, Jain, and
  Wilkinson}]{AvramovicCingara.2009}
Avramovic-Cingara G, Ososkov Y, Jain MK, Wilkinson DS (2009) Effect of
  martensite distribution on damage behaviour in dp600 dual phase steels.
  Materials Science and Engineering A (516):7--16

\bibitem[{Bai and Wierzbicki(2008)}]{Bai.2008}
Bai Y, Wierzbicki T (2008) A new model of metal plasticity and fracture with
  pressure and lode dependence. International Journal of Plasticity
  24:1071--1096

\bibitem[{Besson(2010)}]{Besson.2010}
Besson J (2010) Continuum models of ductile fracture: A review. International
  Journal of Damage Mechanics 19(1):3--52, \doi{10.1177/1056789509103482}

\bibitem[{Bieler et~al.(2009)Bieler, Eisenlohr, Roters, Kumar, Mason, Crimp,
  and Raabe}]{Bieler.2009}
Bieler TR, Eisenlohr P, Roters F, Kumar D, Mason DE, Crimp MA, Raabe D (2009)
  The role of heterogeneous deformation on damage nucleation at grain
  boundaries in single phase metals. International Journal of Plasticity
  25(9):1655--1683, \doi{10.1016/j.ijplas.2008.09.002}, pII: S074964190800140X

\bibitem[{B{\o}rvik et~al.(2001)B{\o}rvik, Hopperstad, Berstad, and
  Langseth}]{Brvik.2001}
B{\o}rvik T, Hopperstad O, Berstad T, Langseth M (2001) A computational model
  of viscoplasticity and ductile damage for impact and penetration. European
  Journal of Mechanics - A/Solids 20(5):685--712,
  \doi{10.1016/S0997-7538(01)01157-3}

\bibitem[{Bouchard et~al.(2011)Bouchard, Bourgeon, Fayolle, and
  Mocellin}]{Bouchard.2011}
Bouchard PO, Bourgeon L, Fayolle S, Mocellin K (2011) An enhanced lemaitre
  model formulation for materials processing damage computation. International
  Journal of Material Forming 4(3):299--315, \doi{10.1007/s12289-010-0996-5}

\bibitem[{Davies and Magee(1979)}]{Davies.1979}
Davies RG, Magee CL (1979) Physical metallurgy of automotive high-strength
  steels. Journal of Metals 31(11):17--23, \doi{10.1007/BF03354565}, pII:
  BF03354565

\bibitem[{Ghadbeigi et~al.(2010)Ghadbeigi, Pinna, Celotto, and
  Yates}]{Ghadbeigi.2010}
Ghadbeigi H, Pinna C, Celotto S, Yates JR (2010) Local plastic strain evolution
  in a high strength dual-phase steel. Materials Science and Engineering A
  (527):5026--5032

\bibitem[{Ghadbeigi et~al.(2013)Ghadbeigi, Pinna, and Celotto}]{Ghadbeigi.2013}
Ghadbeigi H, Pinna C, Celotto S (2013) Failure mechanisms in dp600 steel:
  Initiation, evolution and fracture. Materials Science and Engineering: A
  588:420--431, \doi{10.1016/j.msea.2013.09.048}, pII: S0921509313010241

\bibitem[{Gurson(1977)}]{Gurson.1977}
Gurson AL (1977) Continuum theory of ductile rupture by void nucleation and
  growth: Part i---yield criteria and flow rules for porous ductile media.
  Journal of Engineering Materials and Technology 99(1):2,
  \doi{10.1115/1.3443401}

\bibitem[{He et~al.(1984)He, Terao, and Berghezan}]{He.1984}
He XJ, Terao N, Berghezan A (1984) Influence of martensite morphology and its
  dispersion on mechanical properties and fracture mechanisms of fe-mn-c dual
  phase steels. Metal Science 18(7):367--373, \doi{10.1179/030634584790419953},
  doi: 10.1179/030634584790419953 doi: 10.1179/030634584790419953

\bibitem[{Hering et~al.(2019)Hering, Dunlap, Tekkaya, Aretz, and
  Schwedt}]{Hering.2019}
Hering O, Dunlap A, Tekkaya A, Aretz A, Schwedt A (2019) Characterization of
  damage in forward rod extruded parts. International Journal of Material
  Forming \doi{10.1007/s12289-019-01525-z}

\bibitem[{Johnson and Cook(1985)}]{Johnson.1985}
Johnson GR, Cook WH (1985) Fracture characteristics of three metals subjected
  to various strains, strain rates, temperatures and pressures. Engineering
  Fracture Mechanics 21(1):31--48, \doi{10.1016/0013-7944(85)90052-9}

\bibitem[{Kadkhodapour et~al.(2011)Kadkhodapour, Butz, and {Ziaei
  Rad}}]{Kadkhodapour.2011}
Kadkhodapour J, Butz A, {Ziaei Rad} S (2011) Mechanisms of void formation
  during tensile testing in a commercial, dual-phase steel. Acta Materialia
  59:2575--2588

\bibitem[{Keim et~al.(2019)Keim, Nonn, and M{\"u}nstermann}]{Keim.2019}
Keim V, Nonn A, M{\"u}nstermann S (2019) Application of the modified
  bai-wierzbicki model for the prediction of ductile fracture in pipelines.
  International Journal of Pressure Vessels and Piping 171:104--116,
  \doi{10.1016/j.ijpvp.2019.02.010}

\bibitem[{Lemaitre(1985)}]{Lemaitre.1985}
Lemaitre J (1985) A continuous damage mechanics model for ductile fracture.
  Journal of Engineering Materials and Technology (107):83--89

\bibitem[{Lemaitre(1992)}]{Lemaitre.1992}
Lemaitre J (1992) A Course on Damage Mechanics. {Springer Verlag}

\bibitem[{Lian et~al.(2013)Lian, Sharaf, Archie, and
  M{\"u}nstermann}]{Lian.2012}
Lian J, Sharaf M, Archie F, M{\"u}nstermann S (2013) A hybrid approach for
  modelling of plasticity and failure behaviour of advanced high-strength steel
  sheets. International Journal of Damage Mechanics 22(2):188--218

\bibitem[{Lian et~al.(2014)Lian, Yang, Vajragupta, M{\"u}nstermann, and
  Bleck}]{Lian.2014}
Lian J, Yang H, Vajragupta N, M{\"u}nstermann S, Bleck W (2014) A method to
  quantitatively upscale the damage initiation of dual-phasesteels under
  various stress states from microscale to macroscale. Computational Materials
  Science 94:245--257

\bibitem[{Liu et~al.(2020)Liu, Lian, M{\"u}nstermann, Zeng, and
  Fang}]{Liu.2020}
Liu W, Lian J, M{\"u}nstermann S, Zeng C, Fang X (2020) Prediction of crack
  formation in the progressive folding of square tubes during dynamic axial
  crushing. International Journal of Mechanical Sciences 176:105534,
  \doi{10.1016/j.ijmecsci.2020.105534}

\bibitem[{Maire et~al.(2008)Maire, Bouaziz, Di~Michiel, and Verdu}]{Maire.2008}
Maire E, Bouaziz O, Di~Michiel M, Verdu C (2008) Initiation and growth of
  damage in a dual-phase steel observed by x-ray microtomography. Acta
  Materialia 56:4954--4964

\bibitem[{Marteau et~al.(2013)Marteau, Haddadi, and Bouvier}]{Marteau.2013}
Marteau J, Haddadi H, Bouvier S (2013) Investigation of strain heterogeneities
  between grains in ferritic and ferritic-martensitic steels. Experimental
  Mechanics 53(3):427--439, \doi{10.1007/s11340-012-9657-6}, pII: 9657

\bibitem[{Meya et~al.(2019)Meya, Kusche, L{\"o}bbe, Al-Samman, Korte-Kerzel,
  and Tekkaya}]{Meya.2019}
Meya R, Kusche C, L{\"o}bbe C, Al-Samman T, Korte-Kerzel S, Tekkaya A (2019)
  Global and high-resolution damage quantification in dual-phase steel bending
  samples with varying stress states. Metals 9, \doi{10.3390/met9030319}

\bibitem[{Mu et~al.(2020)Mu, Jia, Ma, Shen, Sun, and Zang}]{Mu.2020}
Mu L, Jia Z, Ma Z, Shen F, Sun Y, Zang Y (2020) A theoretical prediction
  framework for the construction of a fracture forming limit curve accounting
  for fracture pattern transition. International Journal of Plasticity p
  102706, \doi{10.1016/j.ijplas.2020.102706}

\bibitem[{M{\"u}nstermann et~al.(2017)M{\"u}nstermann, Lian, P{\"u}tz,
  K{\"o}nemann, and Brinnel}]{Munstermann.2017}
M{\"u}nstermann S, Lian J, P{\"u}tz F, K{\"o}nemann M, Brinnel V (2017)
  Comparative study on damage evolution during sheet metal forming of steels
  dp600 and dp1000. Journal of Physics: Conference Series 896

\bibitem[{Novokshanov et~al.(2015)Novokshanov, D{\"o}bereiner, Sharaf, and
  M{\"u}nstermann}]{Novokshanov.2015}
Novokshanov D, D{\"o}bereiner B, Sharaf M, M{\"u}nstermann S (2015) A new model
  for upper shelf impact toughness assessment with a computationally efficient
  parameter identification algorithm. Engineering Fracture Mechanics
  148:281--303

\bibitem[{Papasidero et~al.(2015)Papasidero, Doquet, and
  Mohr}]{Papasidero.2015}
Papasidero J, Doquet V, Mohr D (2015) Ductile fracture of aluminum 2024-t351
  under proportional and non-proportional multi-axial loading: Bao--wierzbicki
  results revisited. International Journal of Solids and Structures
  69-70:459--474, \doi{10.1016/j.ijsolstr.2015.05.006}, pII: S0020768315002164

\bibitem[{Park et~al.(2014)Park, Nishiyama, Nakada, Tsuchiyama, and
  Takaki}]{Park.2014}
Park K, Nishiyama M, Nakada N, Tsuchiyama T, Takaki S (2014) Effect of the
  martensite distribution on the strain hardening and ductile fracture
  behaviors in dual-phase steel. Materials Science and Engineering: A
  604:135--141, \doi{10.1016/j.msea.2014.02.058}, pII: S0921509314002111

\bibitem[{Ramazani et~al.(2013)Ramazani, Schwedt, Aretz, Prahl, and
  Bleck}]{Ramazani.2013}
Ramazani A, Schwedt A, Aretz A, Prahl U, Bleck W (2013) Characterization and
  modelling of failure initiation in dp~steel. Computational Materials Science
  (75):35--44

\bibitem[{Rueden et~al.(2017)Rueden, Schindelin, Hiner, DeZonia, Walter, Arena,
  and Eliceiri}]{Rueden.2017}
Rueden CT, Schindelin J, Hiner MC, DeZonia BE, Walter AE, Arena ET, Eliceiri KW
  (2017) Imagej2: Imagej for the next generation of scientific image data. BMC
  bioinformatics 18(1):529, \doi{10.1186/s12859-017-1934-z}, journal Article,
  \eprint{29187165}

\bibitem[{Saai et~al.(2014)Saai, Hopperstad, Granbom, and Lademo}]{Saai.2014}
Saai A, Hopperstad OS, Granbom Y, Lademo OG (2014) Influence of volume fraction
  and distribution of martensite phase on the strain localization in dual phase
  steels. Procedia Materials Science 3:900--905,
  \doi{10.1016/j.mspro.2014.06.146}, pII: S2211812814001473

\bibitem[{Schindelin et~al.(2012)Schindelin, Arganda-Carreras, Frise, Kaynig,
  Longair, Pietzsch, Preibisch, Rueden, Saalfeld, Schmid, Tinevez, White,
  Hartenstein, Eliceiri, Tomancak, and Cardona}]{Schindelin.2012}
Schindelin J, Arganda-Carreras I, Frise E, Kaynig V, Longair M, Pietzsch T,
  Preibisch S, Rueden C, Saalfeld S, Schmid B, Tinevez JY, White DJ,
  Hartenstein V, Eliceiri K, Tomancak P, Cardona A (2012) Fiji: an open-source
  platform for biological-image analysis. Nature methods 9(7):676--682,
  \doi{10.1038/nmeth.2019}, journal Article Research Support, N.I.H.,
  Extramural Research Support, Non-U.S. Gov't, \eprint{22743772}

\bibitem[{Schowtjak et~al.(2019)Schowtjak, Wang, Hering, Clausmeyer, Lohmar,
  Schulte, Ostwald, Hirt, and Tekkaya}]{Schowtjak.2019}
Schowtjak A, Wang S, Hering O, Clausmeyer T, Lohmar J, Schulte R, Ostwald R,
  Hirt G, Tekkaya A (2019) Prediction and analysis of damage evolution during
  caliber rolling and subsequent cold forward extrusion. Production Engineering
  \doi{10.1007/s11740-019-00935-x}

\bibitem[{Shen et~al.(2020)Shen, M{\"u}nstermann, and Lian}]{Shen.2020}
Shen F, M{\"u}nstermann S, Lian J (2020) Investigation on the ductile fracture
  of high-strength pipeline steels using a partial anisotropic damage mechanics
  model. Engineering Fracture Mechanics 227:106900,
  \doi{10.1016/j.engfracmech.2020.106900}

\bibitem[{Shen et~al.(1986)Shen, Lei, and Liu}]{Shen.1986}
Shen HP, Lei TC, Liu JZ (1986) Microscopic deformation behaviour of
  martensitic--ferritic dual-phase steels. Materials Science and Technology
  2(1):28--33, \doi{10.1179/mst.1986.2.1.28}, doi: 10.1179/mst.1986.2.1.28 doi:
  10.1179/mst.1986.2.1.28

\bibitem[{Tasan et~al.(2010)Tasan, Hoefnagels, and Geers}]{Tasan.2010}
Tasan CC, Hoefnagels J, Geers M (2010) Microstructural banding effects
  clarified through micrographic digital image correlation. Scripta Materialia
  62(11):835--838, \doi{10.1016/j.scriptamat.2010.02.014}

\bibitem[{Tasan et~al.(2014{\natexlab{a}})Tasan, Diehl, Yan, Zambaldi,
  Shanthraj, Roters, and Raabe}]{Tasan.2014}
Tasan CC, Diehl M, Yan D, Zambaldi C, Shanthraj P, Roters F, Raabe D
  (2014{\natexlab{a}}) Integrated experimental--simulation analysis of stress
  and strain partitioning in multiphase alloys. Acta Materialia 81:386--400,
  \doi{10.1016/j.actamat.2014.07.071}, pII: S1359645414005898

\bibitem[{Tasan et~al.(2014{\natexlab{b}})Tasan, Hoefnagels, Diehl, Yan,
  Roters, and Raabe}]{Tasan.2014b}
Tasan CC, Hoefnagels J, Diehl M, Yan D, Roters F, Raabe D (2014{\natexlab{b}})
  Strain localization and damage in dual phase steels investigated by coupled
  in-situ deformation experiments and crystal plasticity simulations.
  International Journal of Plasticity 63:198--210,
  \doi{10.1016/j.ijplas.2014.06.004}

\bibitem[{Tasan et~al.(2015)Tasan, Diehl, Yan, Bechtold, Roters, Schemmann,
  Zheng, Peranio, Ponge, Koyama, Tsuzaki, and Raabe}]{Tasan.2015}
Tasan CC, Diehl M, Yan D, Bechtold M, Roters F, Schemmann L, Zheng C, Peranio
  N, Ponge D, Koyama M, Tsuzaki K, Raabe D (2015) An overview of dual-phase
  steels: Advances in microstructure-oriented processing and micromechanically
  guided design. Annual Review of Materials Research 45(1):391--431,
  \doi{10.1146/annurev-matsci-070214-021103}

\bibitem[{Tekkaya et~al.(2017)Tekkaya, {Ben Khalifa}, Hering, Meya, Myslicki,
  and Walther}]{Tekkaya.2017}
Tekkaya AE, {Ben Khalifa} N, Hering O, Meya R, Myslicki S, Walther F (2017)
  Forming-induced damage and its effects on product properties. CIRP Annals
  66(1):281--284, \doi{10.1016/j.cirp.2017.04.113}

\bibitem[{Tvergaard(1981)}]{Tvergaard.1981}
Tvergaard V (1981) Influence of voids on shear band instabilities under plane
  strain conditions. International Journal of Fracture 17(4):389--407,
  \doi{10.1007/BF00036191}

\bibitem[{Tvergaard and Needleman(1984)}]{Tvergaard.1984}
Tvergaard V, Needleman A (1984) Analysis of the cup-cone fracture in a round
  tensile bar. Acta Metallurgica 32(1):157--169,
  \doi{10.1016/0001-6160(84)90213-X}

\bibitem[{West et~al.(2012)West, Lian, M{\"u}nstermann, and Bleck}]{West.2012}
West O, Lian J, M{\"u}nstermann S, Bleck W (2012) Numerical determination of
  the damage parameters of a dual-phase sheet steel. ISIJ International
  52(4):743--752, \doi{10.2355/isijinternational.52.743}

\bibitem[{Wu et~al.(2017)Wu, Li, Di, Brinnel, Lian, and
  M{\"u}nstermann}]{Wu.2017}
Wu B, Li X, Di Y, Brinnel V, Lian J, M{\"u}nstermann S (2017) Extension of the
  modified bai--wierzbicki model for predicting ductile fracture under complex
  loading conditions. Fatigue {\&} Fracture of Engineering Materials {\&}
  Structures 40(12):2152--2168

\end{thebibliography}

%
%

\end{document}